\documentclass[prl,showpacs,twocolumn,amsmath,amssymb,floatfix,superscriptaddress]{revtex4-2}
\usepackage{graphicx} 
\usepackage{bm}
\usepackage{bbm}
\usepackage{color}
\usepackage{amsmath}
\usepackage{amssymb}
\usepackage{color}
\usepackage{footnote}
\usepackage{comment}
\usepackage{hyperref}
\usepackage{subfigure}
\usepackage{appendix}
\usepackage[applemac]{inputenc}

\begin{document}

\title{Universal transverse diode effects in superconducting unconventional magnet hybrids}

\author{Pei-Hao Fu}
\affiliation{Department of Physics and Astronomy, University of Florence, I-50019 Sesto Fiorentino, Italy}

\author{Luca Chirolli}
\email{luca.chirolli@unifi.it}
\affiliation{Department of Physics and Astronomy, University of Florence, I-50019 Sesto Fiorentino, Italy}
 
\author{Jorge Cayao}
\email{jorge.cayao@physics.uu.se}
\affiliation{Department of Physics and Astronomy, Uppsala University, Box 516, S-751 20 Uppsala, Sweden}
 
\date{\today} 
\begin{abstract}
We demonstrate that superconducting unconventional magnet junctions harbor transverse spin and charge diode effects with perfect rectification efficiencies.
This nonreciprocal transverse transport is governed by the interplay of mirror reflections and spin rotation with respect to the N\'{e}el vector, a mechanism that is universally applicable to all unconventional magnets. 
Interestingly, the transverse spin diode features pure spin currents, while the charge diode yields spin-polarized charge currents, both exhibiting highly controllable rectification functionalities at vanishing net magnetization. 
Our results position superconducting unconventional magnets at the forefront of orthogonal engineering in next-generation transverse superconducting spintronics.
 \end{abstract}
\maketitle

The discovery of unconventional magnetism offers a fertile foundation for designing novel superconducting phases and technological applications \cite{Yuri2025Superconducting}.
Crucially, unconventional magnets (UMs) exhibit an intrinsic, nonrelativistic, anisotropic spin splitting  \cite{noda2016momentum, NakaNatCommun2019, Hayami19, Ahn2019, Yuanprb20, LiborSAv, NakaPRB2020, Yuanprm21, LiborPRX22, landscape22, MazinPRX22}, which generates anisotropic spin-polarized Fermi surfaces without net magnetization and vanishing stray fields \cite{Bai2024,MazinPRX22,Song2025,Yuri2025Superconducting}.
Consequently, UMs serve as a cornerstone for several superconducting phenomena, ranging from spin-triplet Cooper pairs \cite{Maeda2025Classification, Chakraborty2024Constraints, khodas2025strain,PhysRevB.111.054520, parshukov2025, mazin2025notes, fu2025light, fu2025floquet, Yokoyama25floquet, Mukasa2025FiniteMomentum, heinsdorf2025, monkman2025perscurrent}, the fundamental building blocks of unconventional superconductivity \cite{Sigrist1991Phenomenological}, to anomalous Josephson effects \cite{zhang2024, Ouassou23, Beenakker23, Bo2024, sun2024, Cheng24, fukaya2024, VosoughiNia2025Altermon, Pal2025Josephson} and nonreciprocal transport \cite{Banerjee2024a, chengdiode24, Chakraborty25, sharma2025diode, Sim2024PairDensity, Ruthvik2025FieldFree} of highly applied interest.
In this regard, nonreciprocal transport is particularly important, as it underpins the realization of quantum circuit elements exhibiting directionally dependent current flow, namely superconducting diodes \cite{Nagaosa2024Nonreciprocal, Tokura2018Nonreciprocal, Shaffer2025Theories, Nadeem2023Superconducting, Wakatsuki2017Nonreciprocal}. 
Despite intense efforts, superconducting diodes in UMs have so far been studied exclusively along the longitudinal direction \cite{Banerjee2024a,  chengdiode24, Chakraborty25, sharma2025diode, Sim2024PairDensity}, leaving transverse diode effects seldom studied.

\begin{figure}[t]
\centering \includegraphics[width=0.45\textwidth]{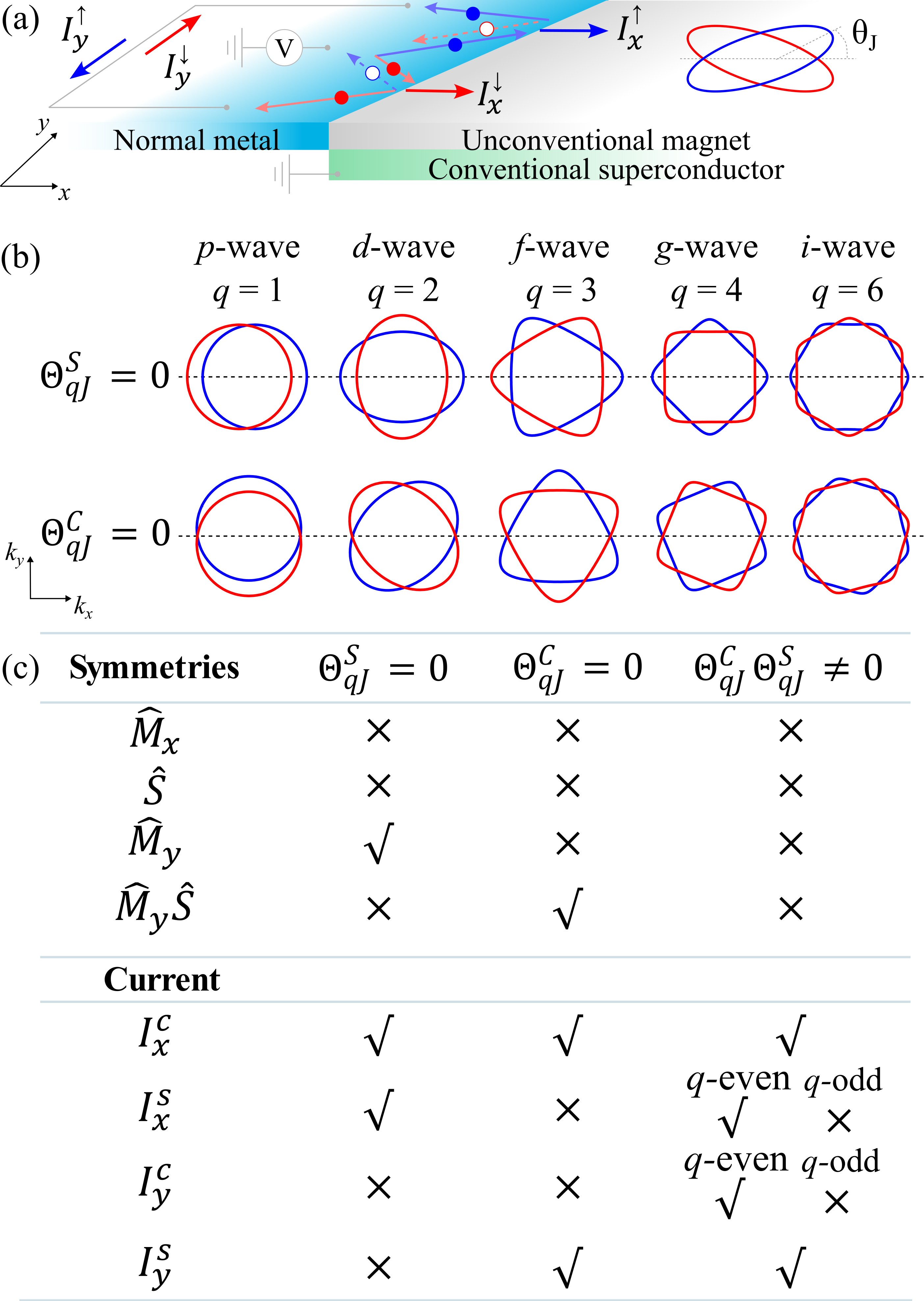}
\caption{
(a) Schematic junction along the $x$-direction formed by a normal metal (N) and an unconventional magnet (UM) with proximately induced $s$-wave superconductivity.
In N, blue (red) filled circles denote spin-up (-down) electrons with propagating direction indicated by solid arrows, while their hole counterparts are denoted by hollow circles and dashed arrows.
In UM, the blue (red) ellipse denotes the spin-up (-down) Fermi surface of a representative $d$-wave UM, with $\theta_J$ denoting the relative orientation between the junction direction ($x$ axis) and the maximum spin splitting.   
(b) Schematic spin-split Fermi surfaces of UMs [Eq.\,(\ref{eq_mkq})] with different parities along the maximum spin-splitting direction ($\Theta^{S}_{qJ}=0$) and the spin-degenerate direction ($\Theta^{C}_{qJ}=0$).
(c) Symmetry constraints and the resulting currents for different $\Theta^{S/C}_{qJ}$. 
The preservation (breaking) of symmetry and current component are indicated by $\checkmark$ ($\times$). 
Longitudinal/transverse charge (spin) currents are denoted as $I_{x/y}^{\text{c(s)}}$.
}
\label{figure1}
\end{figure}

In this work, we consider the recently discovered UMs with proximity-induced spin-singlet $s$-wave superconductivity [Fig.\ref{figure1}(a)] and reveal the emergence of the transverse spin and charge diode effects with perfect rectification efficiencies.
Furthermore, the transverse spin diode features pure spin currents, while the charge diode yields spin-polarized charge currents, both exhibiting highly controllable rectification functionalities yet vanishing net magnetization.
More interestingly, both transverse nonreciprocities are governed by the interplay of mirror reflections of the junction structure and spin rotation with respect to the N\'{e}el vector, which thereby provides a universal mechanism applicable to all UMs [Fig.\ref{figure1}(b,c)]. Our findings establish superconducting unconventional magnet junctions as a foundational venue for orthogonal engineering, thereby paving the way for next-generation transverse spintronics at zero net magnetization.

\textit{Symmetries and universality}.---We begin by presenting a unified picture of required symmetries for realizing transverse spin and charge diodes in junctions formed by UMs with conventional spin-singlet $s$-wave superconductivity \footnote{We note that the analysis presented can also be extended to UMs with spin-triplet $p$-wave and spin-singlet $d$-wave superconductivity.}, as shown in Fig.\ref{figure1}(a). 
To this end, we highlight that the key ingredient for transverse diodes stems from the anisotropic spin splitting of UMs. 
Thus, to unveil the symmetry requirements and universality from the perspective of UMs, at this point it is sufficient to analyze the UM field along $z$ given by  $\mathcal{M}_{\bm{k}}^{q}=M_{\bm{k}}^{q}\sigma_{z}$ \cite{Yuri2025Superconducting}, with
\begin{equation}
M_{\bm{k}}^{q} = J (k/k_F)^{q} \cos\!\left(q\theta_k - q\theta_{J}\right),
\label{eq_mkq}
\end{equation}
where $q=\{2,4,6\}$ corresponds to even-parity $d$-, $g$-, and $i$-wave UMs (known as altermagnets), while $q=\{1,3\}$ to odd-parity $p$- and $f$-wave UMs.
In UMs, the spin-rotation symmetry $\hat{S} = \sigma_x$ and the $2q$-fold rotational symmetry $\hat{C}_{2q,z}$ are individually broken, while their combined operation $\hat{S}\hat{C}_{2q,z}$ is preserved. 
The resulting Fermi surfaces are anisotropically spin split with the parity given by $q$, see Fig.\,\ref{figure1}(b).
The above symmetries and anisotropic spin splitting of UMs remain when spin-singlet $s$-wave superconductivity is proximity-induced.
However, when attaching a normal electrode, $\hat{S}\hat{C}_{2q,z}$ is broken in the whole junction [Fig.\,\ref{figure1}(a)], while the anisotropic spin splitting in UMs persists.
Moreover, in this setup, the mirror symmetries  $\hat{M}_{x}$  and  $\hat{M}_{y}$ with respect and perpendicular to the interface become essential since they combine with the symmetries of UMs to enable unique transverse nonreciprocity, as explained below. 

In the considered junction, since the normal lead and the conventional superconductor are both isotropic and spin-degenerate, the spin rotation symmetry $\hat{S}$ and the mirror symmetry $\hat{M}_{x}$ with respect to the junction interface are naturally broken by the spin splitting of UMs.
The anisotropy depends on the orientation of the UM relative to the junction direction, i.e., $\theta_J$ in Eq.\,(\ref{eq_mkq}), hence linked to the presence of $\hat{M}_{y}$ and $\hat{M}_{y}\hat{S}$.
To understand this, we rewrite Eq.\,(\ref{eq_mkq}) as $M_{\bm{k}}^{q} = J \![ P_{c}^{q}(\bm{k}) \Theta_{qJ}^{c} + P_{s}^{q}(\bm{k}) \Theta_{qJ}^{s} ]$, where $\Theta_{qJ}^{c} = \cos(q\theta_J)$, $\Theta_{qJ}^{s} = \sin(q\theta_J)$ and $P_{c(s)}^{q}(\bm{k})$ are polynomials of $k_x$ and $k_y$  satisfying a parity-dependent evenness (oddness) as $P_{c(s)}^{q}(\bm{k}) = (-1)^{q} P_{c(s)}^{q}(-\bm{k})$.
Thus, the symmetry $\hat{M}_{y}\hat{S}$  [$\hat{M}_{y}$] is preserved when $\theta_{J}=(2n+1) \pi /( 2q) $ [$\theta _{J}=n\pi /q$], which corresponds to $\Theta^{c}_{qJ}=0$ [$\Theta^{s}_{qJ}=0$]; by analogy, both symmetries are broken when $\Theta^{c}_{qJ}\Theta^{s}_{qJ}\neq 0$; see  Figs.\,\ref{figure1}(b,c).
As a result, while $\hat{S}$ and $\hat{M}_{x}$ are naturally broken by UMs, $\hat{M}_{y}$ and $\hat{M}_{y}\hat{S}$ are tied to the anisotropic spin splitting with respect to the junction direction, unveiling a unique feature of UM-based junctions [Figs.\,\ref{figure1}(a)].

A remarkable consequence of the above symmetries is that they determine nonreciprocal transport in superconducting junctions based on UMs [Fig.\,\ref{figure1}(a)]. The resulting nonreciprocal currents manifest in longitudinal/transverse charge (c) and spin (s) currents defined as $I_{x/y}^{\text{c(s)}}=I_{x/y}^{\uparrow}\pm I_{x/y}^{\downarrow}$, with  $I_{x/y}^{\uparrow(\downarrow)}$ the spin resolved currents  [Fig.\,\ref{figure1}(a,c)].
When the junction is aligned with the direction of maximal spin splitting ($\Theta^{s}_{qJ}=0$), the breaking of $\hat{S}$ induces a net spin current $I_x^{\text{s}} \neq 0$ and the absence of $\hat{M}_{x}$ causes a nonreciprocity in longitudinal charge and spin currents, i.e., $I_x^{\text{c/s}}(eV) \neq -I_x^{\text{c/s}}(-eV)$; hence, a \textit{longitudinal spin diode} effect arises, explaining the origin of the longitudinal spin diodes in altermagnet-superconductor setups \cite{Fu2026Perfectspin}.
For a junction along the spin-degenerate direction ($\Theta^{c}_{qJ}=0$), breaking $\hat{M}_y$ induces a transverse current due to the imbalance between upward ($y > 0$) and downward ($y < 0$) propagating electrons. However, because the combined  $\hat{M}_y \hat{S}$ symmetry is preserved, the spin resolved currents are constrained to be equal in magnitude but opposite in sign  ( $I_y^{\uparrow}=-I_y^{\downarrow}$), thereby enabling a \textit{transverse spin diode} with pure spin currents ($I_y^{\text{s}}$) and vanishing charge current ($I_y^{\text{c}}$).
Furthermore, when the junctions deviate from the spin-degenerate direction  ($\Theta^{c}_{qJ}\Theta^{s}_{qJ}\neq0$),  
breaking $\hat{M}_y \hat{S}$ enables an imbalance between upward-going and downward-going opposite-spin currents ($I_y^{\uparrow}\neq-I_y^{\downarrow}$), giving rise to finite transverse spin-polarized currents that define the \textit{transverse charge diode effect}.
Therefore, the collective interplay of the spin rotation symmetry $\hat{S}$, together with the combined symmetry $\hat{M}_{y}\hat{S}$ intrinsic to UMs, and mirror symmetries $\hat{M}_{x/y}$, determine nonreciprocal transport and enable the existence of transverse diode effects in UM-based superconducting junctions, as summarized in Fig.\,\ref{figure1}(b,c).

\begin{figure}[t]
\centering \includegraphics[width=0.45\textwidth]{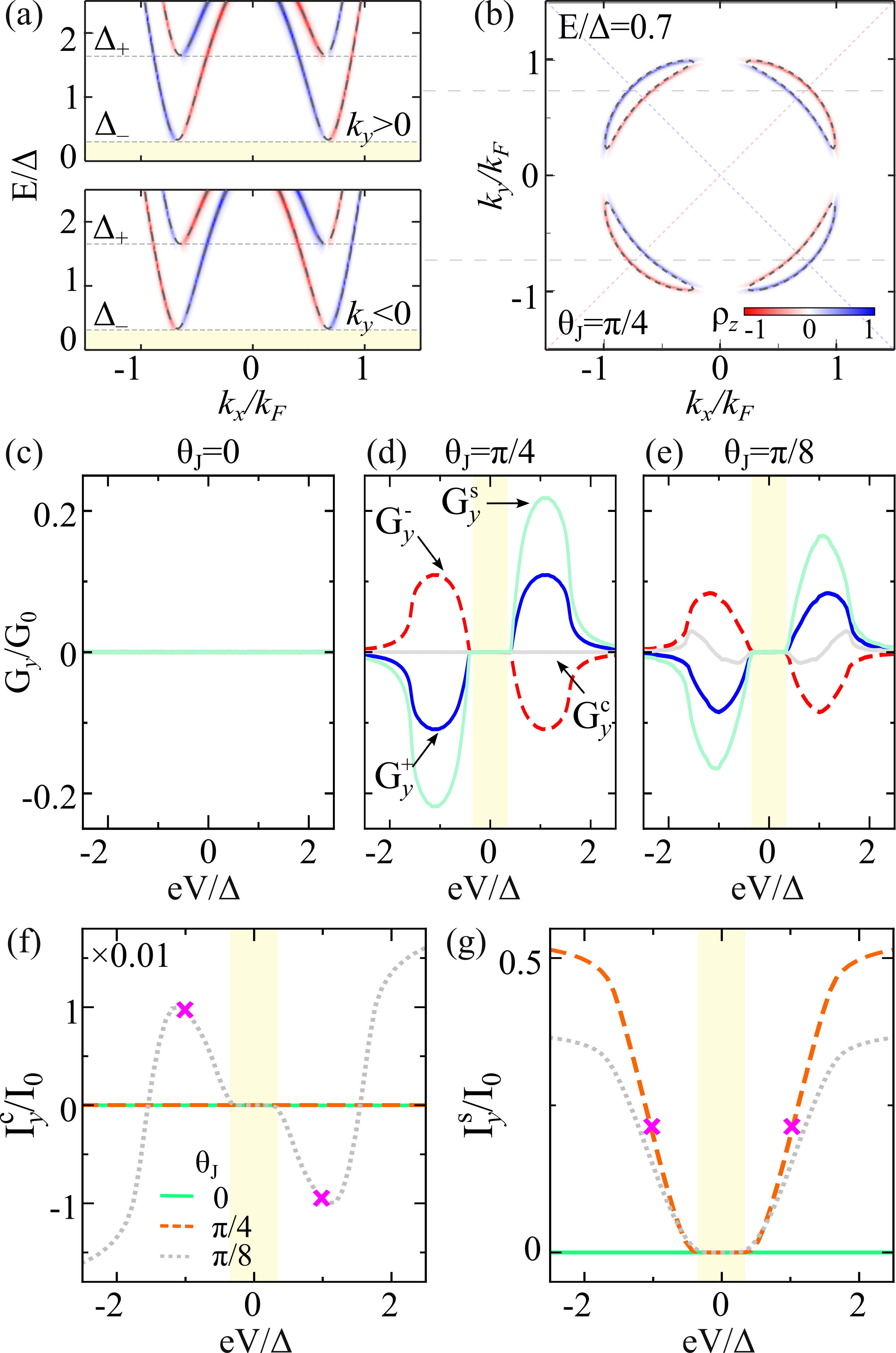}
\caption{
(a) Dispersion and spin density [Eq.\,(\ref{eq_sz})] of $d_{xy}$-wave AM ($\theta_J=\pi/4$) with conventional spin-singlet $s$-wave superconductivity as a function of $k_x$ with $k_y>0$ (upper panel) and $k_y<0$ (lower panel).
The horizontal gray dashed lines denote the critical band edge [Eq.\ (\ref{eq_Deltapm})].
The yellow shaded area marks the fully gaped energy window $|E|<\Delta_-$.
(b) Equal energy contour and spin density at the $k_x$-$k_y$ plane. 
The horizontal dashed line indicates the $k_y$ chosen for (a).
(c-e) Hall conductance as a function of $eV$ with various altermagnetic orientations $\theta_J$.
The blue, red dashed, gray, and light green lines represent the spin-up, spin-down, charge, and spin conductance. respectively. 
(f, g) Charge and spin transverse currents $I_x^{\text{c}}$ and $I_x^{\text{c}}$ as a function of $eV$. 
The magenta crosses mark the currents with $eV=\pm\Delta$. 
Parameters: $\mu = t$, $J=0.1\mu$, $\Delta=0.15\mu$, corresponding to $\Delta_+\approx  1.67\Delta$ and $\Delta_-\approx  0.33\Delta$; junction interface is transparent and $k_{\rm B}T=0.01\Delta$. 
}
\label{figure2}
\end{figure}

\textit{Nonreciprocal transverse transport}.---Having demonstrated the required symmetries and universality of the transverse diode effects, we now investigate the realization of transverse nonreciprocal transport.
For pedagogical purposes, here we focus on a junction formed by a $d$-wave altermagnet (AM) with conventional spin-singlet $s$-wave superconductivity [Fig.\ref{figure1}(a)], which in the basis of $(c_{k,\uparrow}, c_{k,\downarrow},c^\dag_{-k,\uparrow}, c^\dag_{-k,\downarrow})^{\rm T}$ is modelled by \cite{Yuri2025Superconducting}
\begin{equation}
\label{eq_hqbdg}
\mathcal{H}_{\bm{k}} = \xi_{\bm{k}} \tau_z - \Delta\, \sigma_y \tau_y + M^{d}_{\bm{k}} \sigma_z \tau_z,
\end{equation}
where $\sigma_{j}$ ($\tau_{j}$) is the $j$-th Pauli  matrix in spin (Nambu) space, $\xi_{\bm{k}}=\hbar^{2}k^{2}/(2m)-\mu$ is the kinetic term with chemical potential $\mu$, $\bm{k} = (k_x, k_y)$ is the crystal momentum, $\Delta$ the spin-singlet $s$-wave pair potential and $M^{d}_{\bm{k}}$ is the AM field given by Eq.\,(\ref{eq_mkq}) for $q=2$.  In Eq.\,(\ref{eq_hqbdg}), the spin-rotation symmetry $\hat{S} = \sigma_x$ and the $4$-fold rotational symmetry $\hat{C}_{4,z}$ are broken, while their combination $\hat{S}\hat{C}_{2q,z}$ is preserved.
The resulting anisotropic spin splitting is seen in the dispersions $E_{\bm{k}}^{\gamma ,\beta} = \gamma M_{\bm{k}} + \beta \sqrt{\xi_{\bm{k}}^{2} + \Delta^{2}}$ [Fig.\,\ref{figure2}(a)] and the corresponding equal energy contours in Fig.\,\ref{figure2}(b), along with the spin density $\rho_z$ (End Matter). 
A precise characterization is obtained from the anisotropic and spin-resolved band edges  $E_{\text{edge}}^{\gamma,\beta}(\theta_{k}) = \gamma M_{k_F}(\theta_k) + \beta \Delta$; here $M_{k_F,\theta_k}\equiv M_{k\rightarrow k_F,\theta_k}$ with $k_F=\sqrt{2m\mu}/\hbar$.
Particularly, at $\theta_k = \theta_J + n\pi/2$, the positive maximized $M_{k_F}(\theta_k)$ gives rise to the upper [lower] bounds of $E_{\text{edge}}^{\gamma,+}(\theta_{k})$ with $\gamma=+$ [$\gamma=-$],   defined as 
\begin{equation}
E_{\text{edge}}^{\gamma,+}(\theta_J + n\pi/2)\equiv \Delta_\gamma = \left| \Delta + \gamma J  \right|\,;
\label{eq_Deltapm}
\end{equation}
on the other hand, the negative maximized $M_{k_F}(\theta_k)$ at $\theta_k = \theta_J + (2n+1)\pi/2$ results in the the lower [upper] bounds of $E_{\text{edge}}^{\gamma,+}(\theta_{k})$ with $\gamma=+$ [$\gamma=-$], which is also given by Eq.\,(\ref{eq_Deltapm}).
These highest (lowest) band edges are shown indicated by the horizontal dashed lines in Figs.\,\ref{figure2}(a) at positive energies; while they reverse in the negative ($\beta=-$) energy branches (not shown here).

The quasiparticle spectrum is, therefore, defined into three distinct energy regimes via Eq.\,(\ref{eq_Deltapm}): 
(i) For $|E| < \Delta_-$, the system is fully gapped without quasiparticle excitations, see the yellow shaded region in Fig.\,\ref{figure2}(a).  
(ii) For $\Delta_- < |E| < \Delta_+$, the gap is spin-split and direction-selective, resulting in quasiparticle states emerging on a pair of equal-energy contours from opposite spin sectors [Fig.\,\ref{figure2}(b)]. 
(iii) For $|E| > \Delta_+$, the system becomes gapless with quasiparticles occupying all momentum directions.  
As a result, $\Delta_\pm$ and  $E_{\text{edge}}^{\gamma,\beta}$ characterize the anisotropic spin splitting in the spectrum due to the interplay between $d$-wave AM and superconductivity. 
As we show below, this governs the emergence of distinct transverse diode effects in superconducting AMs. 


To analyze the transverse transport in the superconducting junction, we discretize Eq.\,(\ref{eq_hqbdg}) into a tight-binding lattice along $x$ and obtain the zero-temperature spin-resolved transverse conductance as \cite{Fu2025Implementation, Blonder1982Transition}
 \begin{eqnarray} 
G^{\gamma}_{y}(eV) &=& G_{0} \sum_{k_{y}} \text{sgn}(k_y) \!( 1 + R_{\text{N}}^{\gamma} + R_{\text{A}}^{\gamma} ), \label{eq_Gx} 
\end{eqnarray}
where $G_0 = e^2 W / (2\pi\hbar)$  and $W$ is the sample width; $R_{\text{N(A)}}^{\gamma}$ is the $\gamma$-resolved normal (Andreev) reflection probability, obtained via lattice Green's functions \cite{Anantram2008Modeling, LopezSancho1985Nonorthogonal, LopezSancho1984Quick, LopezSancho1985Highly, Fu2022, Fu2022b, Fu2020Transport}.  The factor $\text{sgn}(k_y)$ indicates that electrons moving upward (downward) with $k_y>0$ ($<0$) give positive (negative) Hall conductance \cite{Fu2025Implementation, Ren2014Asymmetric, Ren2013Anomalous, Costa2019Skew, Costa2025Transport, Costa2021Superconducting, Costa2020Anomalous}. 
Hence, both Andreev reflections (ARs) and normal reflections (NRs) contribute to the transverse conductance via injecting Cooper pairs and specular reflections, respectively \cite{Fu2025Implementation, Costa2019Skew}.
Eq.\,(\ref{eq_Gx}) defines the charge and spin conductances via $G_{y}^{\text{c(s)}} = G_{y}^{+} + (-) G_{y}^{-}$.
Moreover, Eq.\,(\ref{eq_Gx}) allows us to obtain the transverse current  as
\cite{Fu2025Implementation, Blonder1982Transition}
\begin{equation}
I^{\gamma}_{y}(eV) = \frac{1}{e} \int dE\, G^{\gamma}_{y}(E)\left[f_{\text{N}}(E,eV)-f_{\text{S}}(E)\right],
\label{eq_iv}
\end{equation}
where $\gamma=\pm$ labels the spin sector, $f_{\text{N}}(E,eV)=f(E-eV)$ and $f_{\text{S}}(E)=f(E)=[1+e^{E/(k_BT)}]^{-1}$ are the Fermi-Dirac distributions in the N and S regions, respectively.  Thereby, the measurable charge (spin) currents are then calculated via $I_{y}^{\text{c(s)}} = I_{y}^{+}+(-)I_{y}^{-}$.

We now analyze the transverse (Hall) conductance and currents in a transparent junction with a superconducting AM obtained from Eqs.\,(\ref{eq_Gx}) and (\ref{eq_iv}), respectively.
The behaviors of the conductance and current are related to the symmetry requirements shown in Fig.\ref{figure1}(c).
To this end, we investigate three junction directions: 
$\theta_J=0$ with $\Theta_{2J}^{S} = 0$, 
$\theta_J=\pi/4$ with $\Theta_{2J}^{C} = 0$, and $\theta_J=\pi/8$ with $\Theta_{2J}^{C}\Theta_{2J}^{S} \neq 0$.
As shown in Figs.\,\ref{figure2}(c-g), both the conductance and currents are finite only when the bias exceeds the gapped region ($eV>\Delta_{-}$) and demonstrate a strong dependence on the AM orientation $\theta_{\rm J}$.
At $\theta_{\rm J}=0$ [Fig.\,\ref{figure2}(c)],
the preserving $\hat{M}_y$ restrict both NRs and ARs contributions in Eq.\,(\ref{eq_Gx}) are even functions in $k_{y}$, which, multiplied by $\text{sgn}(k_y)$, result in $G_y^{\gamma}=0$. 
For details on the behavior of AR and NR.
As a result, the upward-going ($k_y>0$) current is always compensated by the downward-going ($k_y<0$) one, leading to vanishing transverse current [Fig.\,\ref{figure2}(f,g)].
Interestingly, for $\theta_{\rm J}=\pi/8,\pi/4$, the $k_y$-evenness in ARs and NRs is broken because of the absence of $\hat{M}_y$.
Thereby, within the partially gapped region $\Delta_-<|eV|<\Delta_+$, finite distinct spin-resolved transverse conductances $G_{y}^{\gamma}\neq0$ as well as the charge and spin ones [Fig.\,\ref{figure2}(d,e)]. 
Notably, for $\theta_{\rm J}=\pi/4$, the presence of $\hat{M}_y\hat{S}$ causes a compensation of the spin-resolved current, resulting in vanishing transverse charge current but maximized transverse spin current; while for $\theta_{\rm J}=\pi/8$ simultaneously breaking of both $\hat{M}_y\hat{S}$ and $\hat{M}_y$ enable finite Hall current with different spin components flowing along opposite directions, see Figs.\,\ref{figure2}(f,g).
Remarkably, both the transverse spin and charge currents exhibit pronounced nonreciprocity with respect to the bias, i.e., $I_{y}^{\text{c(s)}}(eV)\neq -I_{y}^{\text{c(s)}}(-eV)$, see the magenta crossing in Fig.\,\ref{figure2}(f,g).
Particularly, the transverse spin current is an even function of the bias, $I_{y}^{\text{s}}(eV)=I_{y}^{\text{s}}(-eV)$, meaning that the same spin current flows for opposite bias polarity.

The found transverse transport originates from the intrinsic anisotropic spin-split Fermi surfaces in AMs via $\theta_{\rm J}$ associated with the junction direction. 
In fact, when the bias lies in the partially gapped regime, spin-up electrons incident with $\theta_k\in(0,\pi/2)$ encounter a superconducting gap [Fig.\,\ref{figure2}(a)], and undergo ARs, thereby injecting Cooper pairs and producing a large conductance [Eq.\,(\ref{eq_Gx})].
By contrast, their spin-down counterparts experience a gapless quasiparticle band and are predominantly transmitted or reflected, yielding a much smaller conductance. 
Since $k_y = k_F \sin\theta_k>0$ for $0<\theta_k<\pi/2$, the upward transverse conductance is therefore mainly contributed by spin-up electrons.  
The situation reverses for incident electrons with $\theta_{k}\in(-\pi/2,0)$ [Fig.\,\ref{figure2}(a)], which experience an inverted spin-resolved gap structure and dominate the downward transverse conductance.
As a result, the anisotropic spin-split Fermi surfaces in AMs impose \textit{spin-direction locking} on the transverse conductance depending on $\theta_{\rm J}$. 
Therefore, by rotating the AM orientation $\theta_{\rm J}$, an AM-based NS junction can realize either nonreciprocal transverse spin or charge diode effects.
 

\begin{figure}[t]
\centering \includegraphics[width=0.48\textwidth]{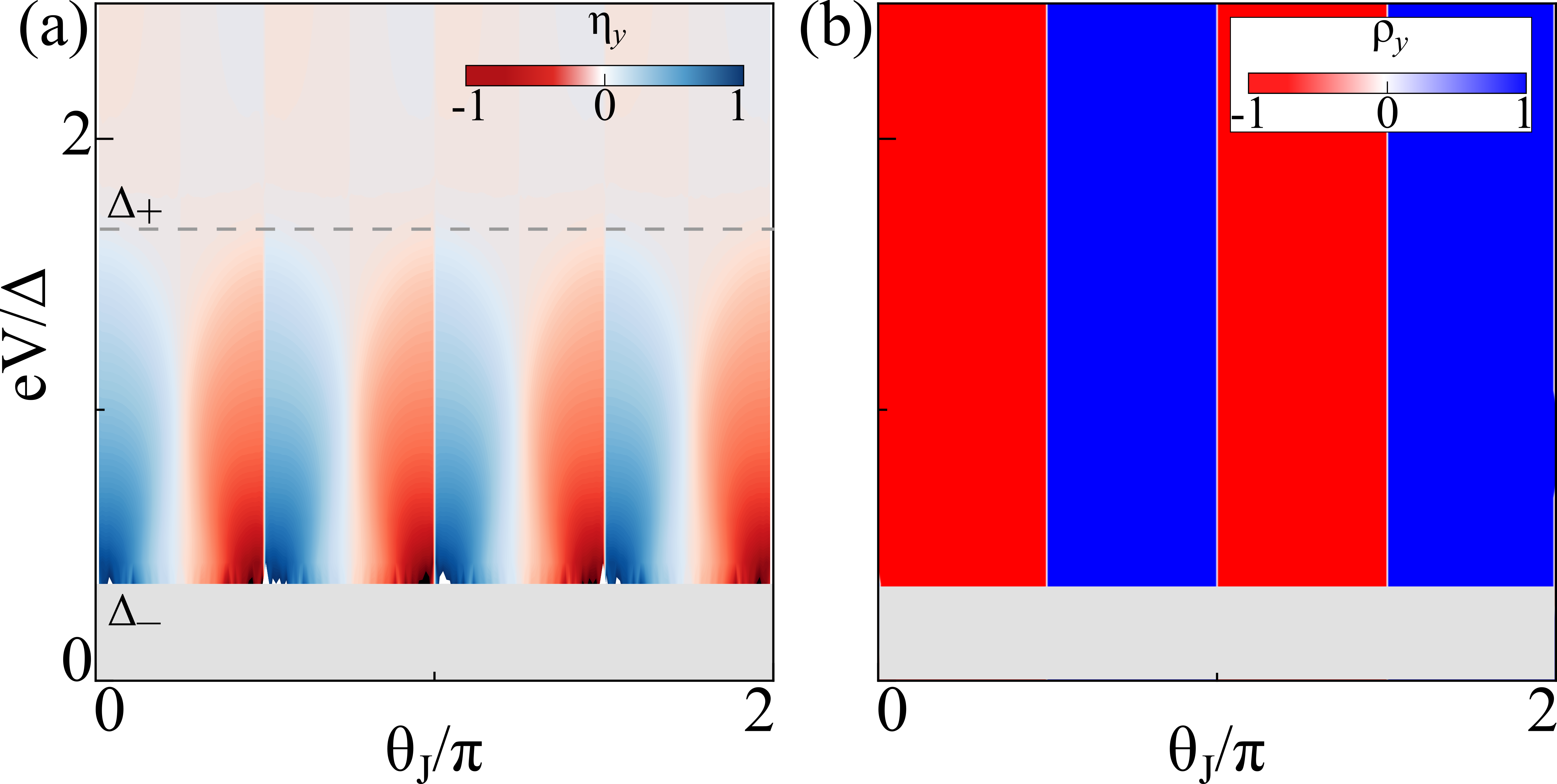}
\caption{
Quality factor of the transverse charge (a) and  spin   diodes (b). The gray regions ($eV<\Delta_-$) indicates that the quality factors are ill-defined due to negligible currents. Parameters: same as Fig.\,\ref{figure2}. 
}
\label{figure3}
\end{figure}

\textit{Quality factor of the transverse diode}.---To quantify the efficiency of the found charge and spin diodes, we now analyze the quality factors associated with the corresponding nonreciprocal currents. 
For the charge currents, we define the quality factor as
\begin{equation}
\eta_{y} = \frac{|I_{+y}| - |I_{-y}|}{|I_{+y}| + |I_{-y}|},
\label{eq_eta}
\end{equation}
where $|I_{\pm y}|$ denotes the magnitude of the current flowing along $\pm y$ direction, and $\text{sgn}(\eta_y)$ specifies the diode's polarity.
The spin-direction locking feature is essential to define the magnitude of the Hall current flowing along $\pm y$ direction: for $0<\theta_J<\pi/4$, the upward (downward) current is carried solely by spin-down (spin-up) quasiparticles, enabling $I_{\pm y}=I_y^{\mp}$; while the spin polarization reverses for $\theta_J\in(\pi/4,\pi/2)$, causing $I_{\pm y}=I_y^{\pm}$.
Thus, the directional spin selectivity of the Hall current enables a $\theta_J$-dependent $I_{\pm y}$, which subsequently affects the charge Hall diode efficiency $\eta_y$ via Eq.\,(\ref{eq_eta}).
Fig.\,\ref{figure3}(a) displays the charge Hall diode efficiency $\eta_y$ as a function of bias $eV$ and the altermagnetic orientation $\theta_J$.  
The Hall diode polarity switches at $\theta_J=n\pi/4$ with vanishing charge Hall current.  
Moreover, $\eta_y$ is ill-defined in the fully gapped region ($eV<\Delta_-$) since $I_{\pm y}=0$ and negligible in the gapless regime ($eV>\Delta_+$) where$I_{+ y} \approx -I_{- y}$.  
By contrast, in the partially gapped region $eV \in (\Delta_-,\Delta_+)$, $\eta_y$ is finite and reaches $\pm1$ as $eV\gtrsim\Delta_-$, characterizing the perfect anomalous Hall diode effect.
We further confirm that the bias and $\theta_J$ windows for the perfect diode effect can be broadened by tuning the interfacial barrier.

The charge diode quality factor defined in Eq.\,(\ref{eq_eta}) does not capture
spin nonreciprocity associated with the spin-direction-locked Hall current:
for example, at $\theta_J=\pi/4$, spin-resolved Hall currents flow in opposite
transverse directions, yielding $\eta_y=0$ while a finite spin Hall effect
persists [Fig.\,\ref{figure2}(e)].  
Thus, to characterize the spin nonreciprocity, it is convenient to first introduce the spin polarization of the current along the $\pm y$ direction, 
\begin{equation}
P_{\pm y}
= \frac{|I_{\pm y}^+| - |I_{\pm y}^{-}|}
{|I_{\pm y}^+| + |I_{\pm y}^{-}|},
\label{eq_Pxy}
\end{equation}
where
$P_{\pm y}>0$ ($<0$) indicates spin-up (spin-down) polarization.  
Then, the degree of spin nonreciprocity is quantified by
\begin{equation}
\rho_y = \left(P_{+y} - P_{-y}\right)/2,
\label{eq_rhoxy}
\end{equation}
such that $\rho_\beta=0$ corresponds to identical spin polarization under current reversal, whereas $\rho_y=\pm1$ represents complete reversal, i.e., a perfect spin diode effect.
Fig.\,\ref{figure3}(b) shows the dependence of $\rho_{y}$ on $\theta_J$ and bias $V$, where perfect spin Hall diode occurs when  $\rho_y=-1$ ($+1$) for $\theta_J \in (0,\pi/4)$ [$\theta_J \in (\pi/4,\pi/2)$] regardless of the bias. 
The perfect spin Hall diode is attributed to the spin-direction-locked transverse current [Figs.\,\ref{figure2}].
The current flowing in the opposite transverse direction is contributed by different spin carriers, yielding $P_{\pm y}=\mp1$ for $\theta_J \in (0,\pi/4)$, which alters for $\theta_J \in (\pi/4,\pi/2)$.
Moreover, since the spin-directional locking effect is  inherited from the intrinsically asymmetric spin-split Fermi surfaces of AMs, the resulting perfect spin Hall diode is insensitive to  interfacial transparency. 
As a result, the emergent ideal spin Hall diode is an intrinsic superconducting nonreciprocal effect that is unique to unconventional magnets.

\begin{figure}
    \centering
    \includegraphics[width=1\linewidth]{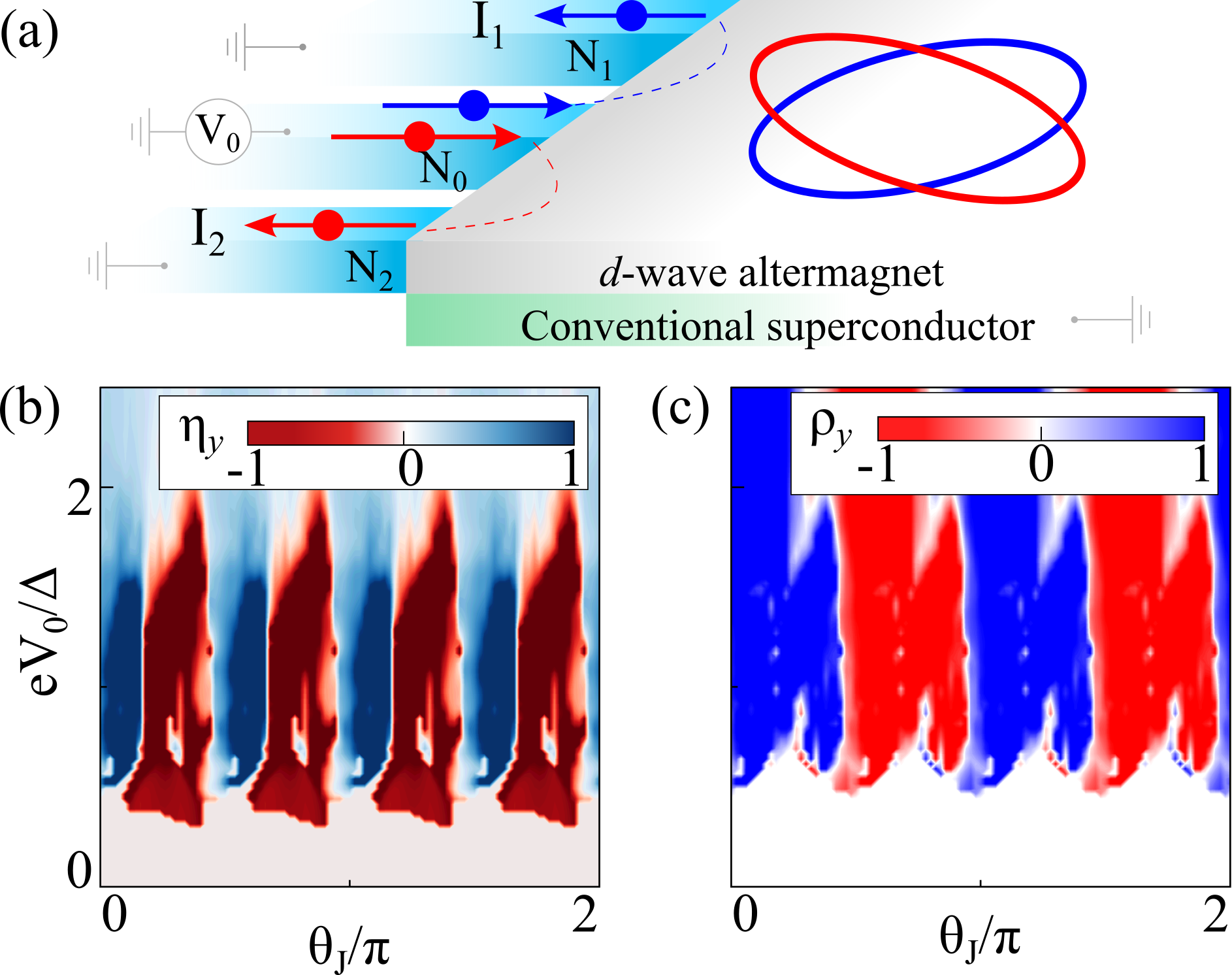}
    \caption{
    (a) Sketch of a four-terminal setup consisting of three normal-metal leads ($N_j$) and one superconducting AM lead, where a bias voltage $V_0$ is applied to the injector lead $N_0$.    The red and blue arrows indicate spin-up (spin-down) transport channels. 
    (b) Charge diode quality factor $\eta_y$ versus   $eV_0$ and      $\theta_J$.
    (c) Same as (b) but for the spin diode quality factor $\rho_y$.
    Parameters: same as Fig.\,\ref{figure2};
    Width of the setup: $30a$; Width of the normal leads $5a$.
}

    \label{figure4}
\end{figure}

\textit{Four-terminal setup.---}
To demonstrate the experimental realization of the proposed transverse diode effects, we consider the four-terminal geometry shown in Fig.\,\ref{figure4}(a), formed by three N leads and one superconducting AM lead.
The central N lead $N_0$ serves as a current injector biased by $V_0$. The transverse transport is revealed in the nonlocal current flowing in $N_{1,2}$  given by \cite{Anantram1996Current,den1996Transport,Pierattelli2025Delta}
\begin{equation}
I_{i}^{\gamma}(eV_{0})
=\frac{1}{e}\int dE\, (T_{e,i}^{\gamma}-T_{h,i}^{\gamma})
\left[f(E,eV_{0})-f(E)\right],
\label{eqIReV}
\end{equation}
where $i\in\{1,2\}$, $T_{e,i}^{\gamma}$ and $T_{h,i}^{\gamma}$ denote the $\gamma$-resolved electron cotunneling and crossed-AR probabilities from $N_0$ to $N_i$, respectively.
The charge (spin) current is defined as
$I_i^{\text{c(s)}}=I_i^{+}+(-)I_i^{-}$.
The  charge diode quality factor is found using Eq.\,(\ref{eq_eta}) with
$I_{+y(-y)}=I_{1(2)}^{\text{c}}$,
while the spin diode is characterized by
$\rho_y=(P_1-P_2)/2$,
where $P_i$ is the current spin polarization   in lead $N_i$ [Eq.\,(\ref{eq_Pxy})].
As illustrated in Fig.\,\ref{figure4}(a), the spin-direction-locking effect induced by the anisotropic spin-split Fermi surfaces remain robust in the multiterminal geometry:
Spin-up (spin-down) electrons injected from $N_0$ are predominantly scattered into the upper (lower) terminal $N_1$ ($N_2$), producing an imbalance of both charge and spin accumulation between the two normal reservoirs.
Consequently, transverse charge and spin nonreciprocity emerge, exhibiting 
$\theta_J$-dependent polarities and nearly perfect rectification with
$\eta_y=\pm1$
and
$\rho_y=\pm1$. Therefore, the proposed transverse diode effects remain robust in a realistic four-terminal geometry and can be directly detected via nonlocal currents collected by side terminals.

In summary, we have uncovered the emergence of transverse spin and charge diode effects with perfect rectification efficiencies in superconducting unconventional magnet hybrids. This joint transverse charge and spin nonreciprocity is strictly governed by the interplay of mirror reflections and spin rotation with respect to the N\'{e}el vector, a mechanism rooted in the intrinsically anisotropic spin splitting that universally characterizes unconventional magnets. Given that longitudinal spin diode behavior has been realized in ferromagnet-superconductor junctions \cite{Strambini2022Superconducting} and quasiparticle-mediated spin Hall effects have been observed in superconductors \cite{Wakamura2015Quasiparticle}, the proposed transverse diode effects  are highly accessible in existing experimental platforms. Promising candidate materials include superconducting hybrids incorporating unconventional magnets such as the $g$-wave altermagnet MnTe \cite{Kazmin2025Andreev} and Co$_{1/4}$NbSe$_2$  \cite{DeVita2025Optical}. By unlocking highly tunable rectification capabilities via pure spin and spin-polarized charge currents at vanishing net magnetization, our findings establish superconducting unconventional magnets as a leading platform for orthogonal engineering in next-generation transverse spintronics.

P.-H. F. and  L. C. acknowledge financial support from the Fondazione Cariplo under the grant 2023-2594.
J. C. acknowledges financial support from the Swedish Research Council (Vetenskapsr{\aa}det Grant No. 2021-04121) and from the Olle Engkvist Foundation (Grant No.  243-1026).  
  
\bibliography{biblio}

\onecolumngrid
\newpage
\begin{center}
\textbf{\textsc{END MATTER}}
\end{center}
\vspace{0.5em}
\twocolumngrid

\textit{\textbf{E1}. Spin density.-}
The spin density of the $d$-wave magnet can be obtained from the Green's function associated with the BdG Hamiltonian [Eq.\,(\ref{eq_hqbdg})], which gives \begin{eqnarray}
\rho_{z}\left( E ,\bm{k}\right) =
-\frac{1}{\pi }\text{Im}%
\text{Tr}[\tau_z \sigma_{z} \mathcal{G}( E+i0^+,\bm{k})] 
\text{,}  \label{eq_sz}
\end{eqnarray}%
where $\mathcal{G}(E+i0^+,\bm{k}) =[(E+i0^+) -\mathcal{H}_{\bm{k}}(\bm{k})]^{-1}$ defines the retarded  Green's function and $0^+$ represents an infinitely small positive number.  Eq.\,(\ref{eq_sz}) shows that the spin density $\rho_{z}(E,\bm{k})$ is directly determined by unconventional magnetism, see Figs.\,\ref{figure2}(a,b).

\begin{figure}[t!]
\centering \includegraphics[width=0.5\textwidth]{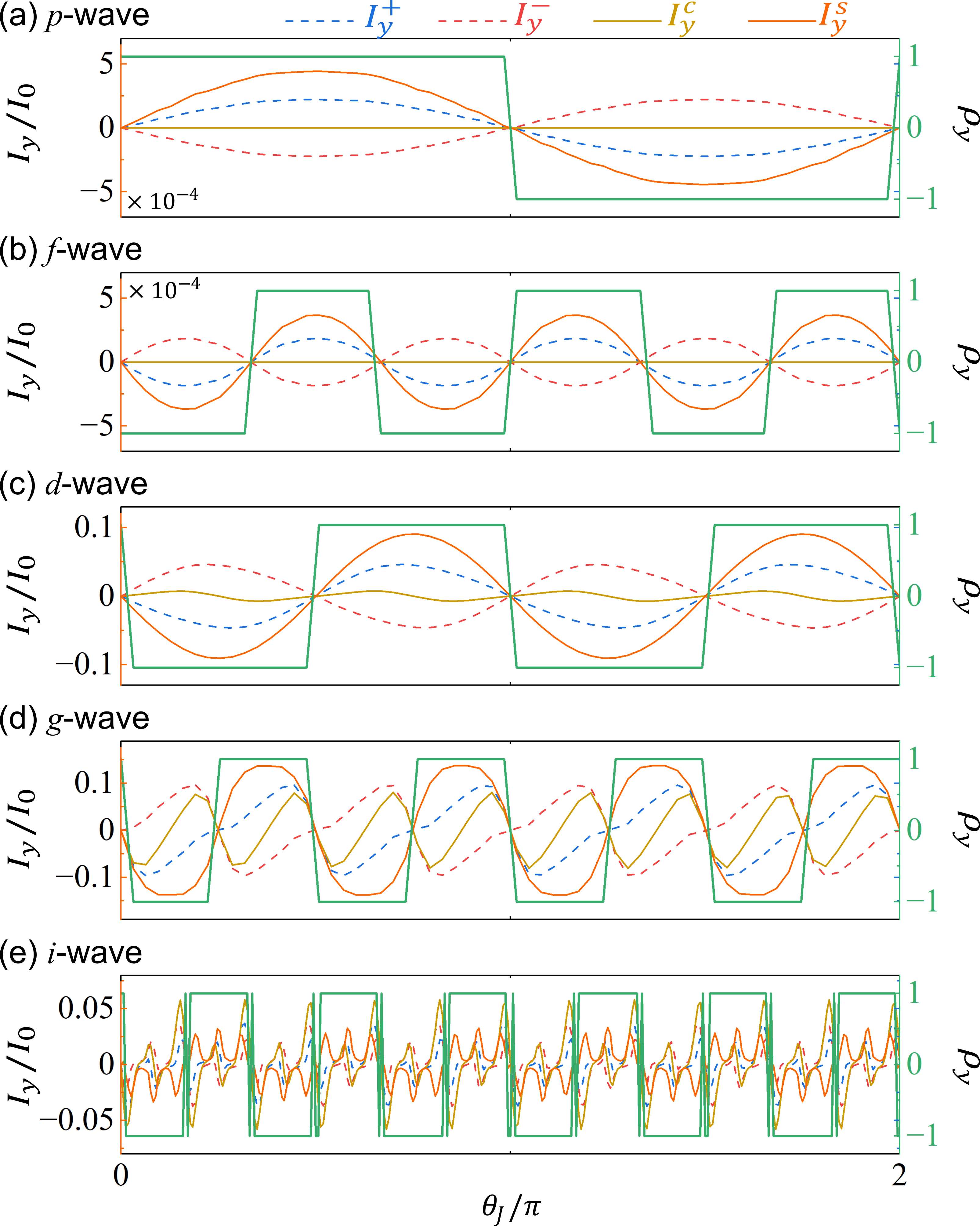}
\caption{
Spin-resolved and charge Hall currents as functions of the UM orientation
$\theta_J$, together with the corresponding spin diode efficiency $\rho_y$.
Parameters are the same as in Figs.\,\ref{figure3} and \ref{figure4}, with
$t_c/t = 1$.
}
\label{figure6}
\end{figure}

\textit{\textbf{E2}. Universality of the diode effect.-} 
The spin and anomalous Hall diode effect are related to the anisotropic spin splitting of the Fermi surface, which is a shared feature of UMs with other parities, and thus are universal for all UM-based NS junctions, as discussed as follows.

Fig.\,\ref{figure6} demonstrates the perfect spin Hall diode effect through the $\theta_{J}$-dependent spin-resolved Hall currents $I_{y}^{\pm}$, and the charge (spin) Hall current $I_{y}^{\text{c(s)}} = I_{y}^{+} + (-) I_{y}^{-}$. 
Results for odd-parity $p$- and $f$-wave UMs with $q = 1$ and $q = 3$ are shown in Figs.\,\ref{figure6}(a,b), while the even-parity $d$-, $g$-, and $i$-wave altermagnets with $q = 2$, $4$, and $6$ are shown in Figs.\,\ref{figure6}(c-e). 
All of spin Hall current exhibits a parity-dependent periodicity of $2\pi/q$. 
For the $p$-, $f$-, $d$-, and $g$-wave UMs [Figs.\,\ref{figure6}(a-d)], the Hall currents clearly follow $
I_{y}^{\pm} \sim J\Theta_{qJ}^{S},
$
and 
$
I_{y}^{\text{s}} \sim J\Theta_{qJ}^{S},
$
respectively. 
For the $i$-wave UM, the relation between $I_{y}^{\text{s}}$ and $\Theta_{6J}^{S}$ is less obvious, but it is still clear that $I_{y}^{\text{s}}=0$ with $\theta_{J}=n\pi/6$, and the charge Hall current exhibits twice the periodicity of the spin Hall current. 
Moreover, the charge Hall current shows an odd-even distinction: while even-parity UMs obey 
$
I_{y}^{\text{c}} \sim J\Theta_{qJ}^{C}\Theta_{qJ}^{S},
$
the odd-parity UMs have $I_{y}^{\text{c}} = 0$. 

The $I_y$-$\theta_{J}$ relation and the even-odd distinction shown in Fig.\,\ref{figure6} reflect the underlying symmetry constraints given by Fig.\,\ref{figure1}(b). 
Because the spin-rotation symmetry $\hat{S}$ and the mirror symmetry $\hat{M}_{x}$ are naturally broken in the presence of UM, the remaining symmetries $\hat{M}_{y}$ and the combined operation $\hat{M}_{y}\hat{S}$ play essential roles in determining transport in the N/S junction for different UM orientations. 
As illustrated in Fig.\,\ref{figure1}, $\hat{M}_{y}\hat{S}$ ($\hat{M}_{y}$) is preserved when 
$\theta_{J} = (2n+1)\pi/(2q) $ $[\theta_{J} = n\pi/q]$,
corresponding to $\Theta_{qJ}^{C}=0$ [$\Theta_{qJ}^{S}=0$], while both symmetries are broken when $\Theta_{qJ}^{C}\Theta_{qJ}^{S}\neq 0$.
Breaking $\hat{M}_{y}$ yields a finite spin-resolved Hall current,  
$I_{y}^{\pm} \sim \Theta_{qJ}^{S}$,  
which vanishes at $\theta_{J}=n\pi/q$.  
However, when $\hat{M}_{y}\hat{S}$ is preserved at  
$\theta_{J} = (2n+1)\pi/(2q)$,  
one has $I_{y}^{+} = -I_{y}^{-}$, resulting in a pure spin Hall effect for UMs of all parities: the spin Hall current is finite while the charge Hall current vanishes. 
In contrast, breaking $\hat{M}_{y}\hat{S}$ produces a finite charge Hall current in even-parity UMs, following  
$I_{y}^{\text{c}} \sim \Theta_{qJ}^{C}\Theta_{qJ}^{S}$ [Figs.\,\ref{figure6}(d-e)].  
Thus, $\hat{M}_{y}\hat{S}$ acts as a pseudo-time-reversal symmetry in even-parity UMs: its preservation protects the spin Hall effect, whereas its breaking yields an anomalous Hall effect with spin-polarized charge Hall currents. 
However, real time-reversal symmetry is preserved for odd-parity $p$- and $f$-wave UMs in Eq.\,(\ref{eq_mkq}); hence, the charge Hall current vanishes for odd $q$, while the spin Hall current persists; see Figs.\,\ref{figure6}(a,b).

\end{document}